\documentclass[%
 reprint,
 amsmath,amssymb,
pre,
]{revtex4-1}

\usepackage{graphicx}
\usepackage{dcolumn}
\usepackage{bm}

\begin{document}


\title{A transient solution for droplet deformation under electric fields}

\author{Jia Zhang,$^1$ Jeffery D. Zahn,$^2$ and Hao Lin$^{1,}$}
\thanks{Email address for correspondence: hlin@jove.rutgers.edu}
\affiliation{$^1$Department of Mechanical and Aerospace Engineering, Rutgers, The State University of New Jersey, Piscataway, NJ 08854, USA \\
$^2$Department of Biomedical Engineering, Rutgers, The State University of New Jersey, Piscataway, NJ 08854, USA}%

\date{\today}

\begin{abstract}
A transient analysis to quantify droplet deformation under DC electric
fields is presented. The full Taylor-Melcher leaky dielectric model
is employed where the charge relaxation time is considered to be finite.
The droplet is assumed to be spheroidal in shape for all times. The
main result is an ODE governing the evolution of the droplet aspect
ratio. The model is validated by extensively comparing predicted deformation
with both previous theoretical and numerical studies, and with experimental
data. Furthermore, the effects of parameters and stresses on deformation
characteristics are systematically analyzed taking advantage of the
explicit formulae on their contributions. The theoretical framework
can be extended to study similar problems, e.g., vesicle electrodeformation and relaxation.
\end{abstract}

\maketitle


\section{Introduction}

When a liquid droplet suspended in another immiscible fluid is subject
to an applied electric field, it undergoes deformation due to the
electrostatic stresses exerted on the interface. Extensive research
on this phenomenon has been conducted to study the deformation due
to its relevance in a variety of industrial applications, including
electrohydrodynamic atomization \cite{Wu2008}, electrohydrodynamic
emulsification \cite{Kanazawa2008}, and ink-jet printing \cite{Basaran2002},
among others. Historically, the deformation dynamics is divided into
two regimes: electrohydrostatics (EHS) and electrohydrodynamics (EHD).
In the first, EHS deformation, the droplet is idealized as a perfect
conductor immersed in a perfect insulating fluid; or both of the fluids
are treated as perfect dielectrics with no free charge \cite{Allan1962,Taylor1964,Miksis1981,Sherwood1988,Ha2000,Dubash2007}. For this case, the electric field only induces
a normal electrostatic stress, which is balanced by surface tension,
and the final equilibrium shape is always prolate. At the steady state,
the hydrodynamic flow is usually absent. In the second, EHD deformation,
both fluids are considered to be leaky dielectrics \cite{Taylor1966,Torza1971,Ajayi1978,Sherwood1988,Vizika1992,Feng1996,Baygents1998,Feng1999,Hirata2000,Ha2000,Bentenitis2005,Sato2006,Lac2007}.
For this case, when an electric field is applied, free charges accumulate
on the droplet surface which induces a tangential electrostatic stress
in addition to the normal one. Driven by this force, the fluids inside
and outside the droplet present toroidal circulations and a viscous
stress is generated in response to balance the tangential electrostatic
stress \cite{Taylor1966}. The droplet deforms into either a prolate
or an oblate spheroid shape depending on the specific electrical properties
of the fluids. With different electrical properties, the effects of
the electrostatic and hydrodynamic stresses on droplet deformation
are distinctive.

This work focuses on a solution method for problems of the second
kind, namely, EHD deformation. This type of problem is more challenging
to solve. In the literature, all theoretical solutions were obtained
largely under two specific assumptions: (i) The deformations are small.
The analysis is performed by assuming that the equilibrium shape of
the droplet only slightly deviates from sphericity. Solutions using
this assumption can be found in \cite{Taylor1966,Torza1971,Ajayi1978}. (ii) For large deformations, the shape is
assumed to be spheroidal during the entire deformation process. Results
using this assumption are given in \cite{Bentenitis2005}. When compared
with experimental data, predictions from the small-deformation theories
always quantitatively underpredict the aspect ratio especially when
the deformation is large. In contrast, the large-deformation theory
has a better agreement both qualitatively and quantitatively. In all
of the above, the theoretical analysis only leads to solutions in
the steady state. The Taylor-Melcher leaky dielectric model \cite{Melcher1969,Saville1997,Zhang2011} with the assumption of
instantaneous charge relaxation has always been used. On the other
hand, the theoretical analysis of transient droplet deformation seems
to attract less attention. Only Dubash and Mestel \cite{Dubash2007} developed a transient
deformation theory for an inviscid, conducting droplet. This analysis,
which solves a EHS deformation problem, is not applicable to study
EHD deformations. In general, to fully solve the transient EHD problem,
numerical simulations have been employed \cite{Baygents1998,Feng1999,Hirata2000,Lac2007}.

In this work, we present a transient analysis of droplet deformation
under direct-current (DC) electric fields. Following Bentenitis and Krause \cite{Bentenitis2005},
we assume the droplet remains spheroidal in shape. The full Taylor-Melcher
leaky dielectric model is employed where the charge relaxation time
is considered finite. In this framework, instantaneous charge relaxation
is treated as a special limiting case. This generalization allows
direct comparison with experimental data which were usually obtained
in fluids with very low conductivities \cite{Ha2000}. The main
result is an ordinary differential equation (ODE) governing the evolution
of the droplet aspect ratio. The availability of this equation allows
us to explicitly analyze the effects of parameters and stresses on
the deformation characteristics. The model is validated by extensively
comparing predicted deformation with both previous theoretical and
numerical studies, and with experimental data. 

\section{Theory}

\begin{figure}
\center

\includegraphics[width=0.35\textwidth]{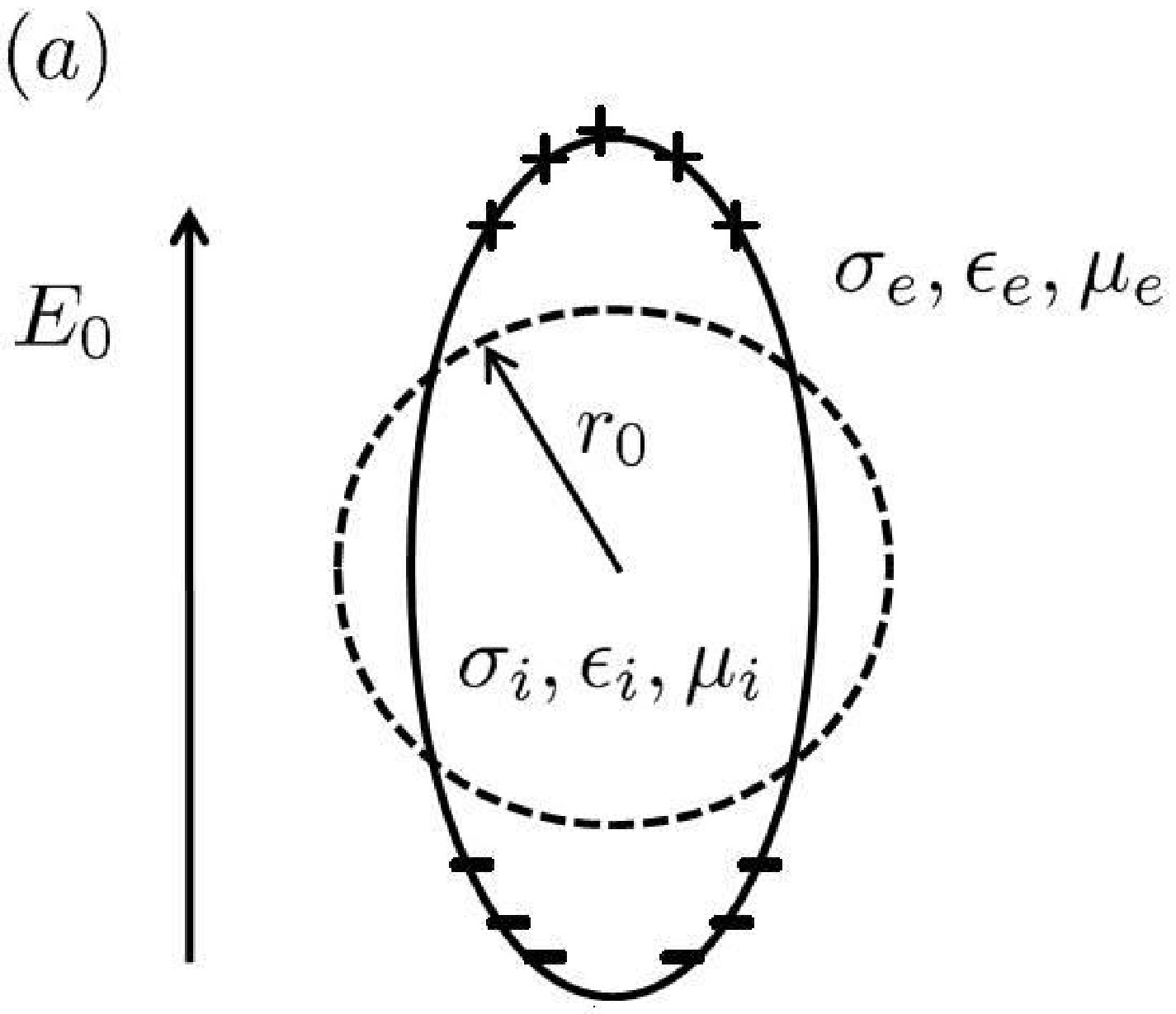}
\includegraphics[width=0.35\textwidth]{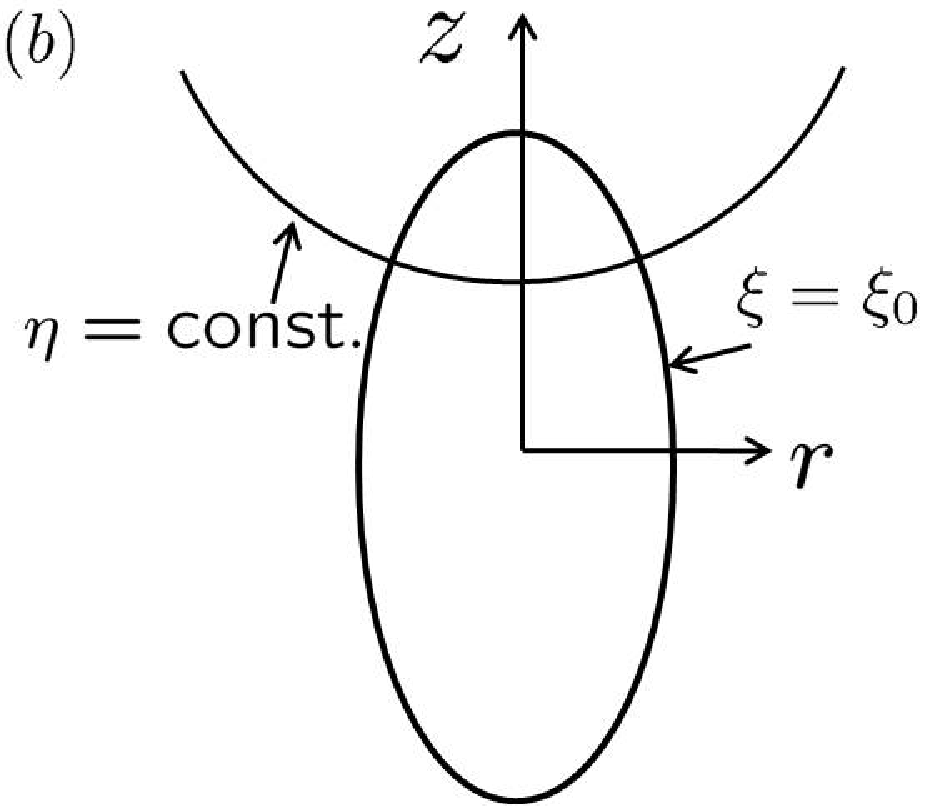}\caption{(a) A schematic of the problem configuration. (b) The prolate spheroidal
coordinate system.\label{fig:Schematics-of-problem.}}

\end{figure}

A schematic of the problem configuration is shown in Fig. \ref{fig:Schematics-of-problem.}(a).
An uncharged, neutrally buoyant liquid droplet of radius $r_{0}$
is suspended in another fluid, and is subject to an applied electric
field of strength $E_{0}$. We assume that the fluids are immiscible
leaky dielectrics with constant electrical and mechanical properties.
$\sigma,\:\epsilon,$ and $\mu$ are the electrical conductivity,
permittivity, and fluid viscosity, and the subscripts $i$ and $e$
denote internal and external, respectively. Under the influence of
an applied electric field, free charges accumulate at the interface,
which induces droplet deformation and EHD flows both inside and outside
the droplet. Taylor \cite{Taylor1966} predicted that droplets may deform
into prolate or oblate shapes depending on the electrical properties
of the fluids. In the following analysis, we focus on developing a
solution for prolate deformations, whereas a solution for oblate deformations
can be pursued in a similar manner (not presented here).

We assume that the droplet remains spheroidal in shape throughout
the process. This approximation is consistent with experimental observations
by Ha and Yang \cite{Ha2000} and Bentenitis and Krause \cite{Bentenitis2005}. Following Tayor \cite{Taylor1964},
Bentenitis and Krause \cite{Bentenitis2005}, and Dubash and Mestel \cite{Dubash2007}, the natural coordinate
system to analyze this problem is the prolate spheroidal coordinate
system, and a schematic is shown in Fig. \ref{fig:Schematics-of-problem.}(b).
The geometry is assumed to be axisymmetric about the $z$ axis, which
aligns with the direction of the applied electric field. The spheroidal
coordinates $(\xi,\:\eta)$ are related to the cylindrical coordinates
$(r,\: z)$ through the equations:\begin{equation}
z=c\xi\eta,\label{spheroidal coordinate z}\end{equation}
\begin{equation}
r=c\sqrt{(\xi^{2}-1)(1-\eta^{2})}.\label{spheroidal coordinate x}\end{equation}
Here $c=\sqrt{a^{2}-b^{2}}$ is chosen to be the semi-focal length
of the spheroidal droplet, and $a$ and $b$ are the major and minor
semi-axis, respectively. The contours for constant $\xi$ are spheroids,
and $\xi\in[1,\:+\infty)$. The contours for constant $\eta$ are
hyperboloids, and $\eta\in[-1,\:1]$. The surface of the prolate spheroid
is conveniently given as\begin{equation}
\xi=\xi_{0}\equiv\frac{a}{c}.\label{prolate surface}\end{equation}
For the derivation below, we further assume that the volume of the
droplet is conserved. We subsequently obtain\begin{equation}
a=r_{0}(1-\xi_{0}^{-2})^{-\frac{1}{3}},\qquad b=r_{0}(1-\xi_{0}^{-2})^{\frac{1}{6}}.\end{equation}
Therefore, the droplet geometry is completely characterized by a single
parameter, $\xi_{0}$, which evolves in time along with deformation.
The critical idea of the current analysis is to express all variables,
e.g., the electric potential and the stream function in terms of $\xi_{0}$. 

In what follows, we will solve the electrical problem first, followed
by a solution of the hydrodynamic problem. An ODE for $\xi_{0}$ is
obtained by applying both the stress matching and kinematic conditions.

\subsection{The electrical problem}

The electric potentials inside and outside the droplet obey the Laplace
equation according to the Ohmic law of current conservation with uniform
electrical conductivity:\begin{equation}
\nabla^{2}\phi_{i}=\nabla^{2}\phi_{e}=0.\label{laplace for droplet}\end{equation}
The matching conditions at the interface are\begin{equation}
||\nabla\phi\cdot\mathbf{\boldsymbol{t}}||=0,\qquad \rm{at}\ \xi=\xi_{0},\label{eq:potential continuous}\end{equation}
\begin{equation}
\frac{\partial q}{\partial t}-||\sigma\nabla\phi\cdot\mathbf{\boldsymbol{n}}||=0,\qquad \rm{at}\ \xi=\xi_{0}.\label{eq:charge conservation}\end{equation}
Here $||\cdot||$ denotes a jump across an interface, and $\mathbf{\boldsymbol{t}}$
and $\mathbf{\boldsymbol{n}}$ are the unit tangential and normal
interfacial vector, respectively. $q=||-\epsilon\nabla\phi\cdot\mathbf{\boldsymbol{n}}||$
is the surface charge density. Note that in Eq. (\ref{eq:charge conservation}),
we have included the displacement current, $\partial q/\partial t$.
This term is particularly important for fluids with very low conductivities
(for example, those used in Ref. \cite{Ha2000}) such that the interfacial
charging time becomes comparable to the deformation time. However,
we have ignored the effect of surface charge convection which is shown
to be small by numerical simulations \cite{Feng1999}. Equation
(\ref{eq:charge conservation}) can be rewritten in terms of the electric
potentials, \begin{eqnarray}
&&\left(\frac{\epsilon_{e}}{c}\frac{\partial\phi_{e}}{\partial\xi}-\frac{\epsilon_{i}}{c}\frac{\partial\phi_{i}}{\partial\xi}\right)\frac{d\frac{c}{h_{\xi}}}{dt}-\left(\frac{\epsilon_{e}}{h_{\xi}c}\frac{\partial\phi_{e}}{\partial\xi}-\frac{\epsilon_{i}}{h_{\xi}c}\frac{\partial\phi_{i}}{\partial\xi}\right)\frac{dc}{dt} \nonumber \\
&&+\left(\frac{\epsilon_{e}}{h_{\xi}}\frac{\partial^{2}\phi_{e}}{\partial\xi\partial t}-\frac{\epsilon_{i}}{h_{\xi}}\frac{\partial^{2}\phi_{i}}{\partial\xi\partial t}\right)+\frac{1}{h_{\xi}}\left(\sigma_{e}\frac{\partial\phi_{e}}{\partial\xi}-\sigma_{i}\frac{\partial\phi_{i}}{\partial\xi}\right)=0, \nonumber \\
&&\qquad\qquad\qquad\qquad\qquad\qquad\qquad\qquad\qquad{\rm at}\ \xi=\xi_{0}.\label{eq:detailed charge conservation}\end{eqnarray}
Here $h_{\xi}$ is a metric coefficient of the prolate spheroidal
coordinate system. The displacement current consists of two parts,
represented by the first three terms on the LHS of the above equation.
The first two terms result from a change in the droplet shape, and the
third results from the charging process as if the shape remains unchanged.
Here, we assume that the first term is negligible compared with the
other two, and Eq. (\ref{eq:detailed charge conservation}) can
be further simplified to be \begin{eqnarray}
\epsilon_{e}\frac{\partial^{2}\phi_{e}}{\partial\xi\partial t}&&-\frac{\epsilon_{e}}{c}\frac{\partial\phi_{e}}{\partial\xi}\frac{dc}{dt}-\epsilon_{i}\frac{\partial^{2}\phi_{i}}{\partial\xi\partial t}+\frac{\epsilon_{i}}{c}\frac{\partial\phi_{i}}{\partial\xi}\frac{dc}{dt} \nonumber\\
&&+\sigma_{e}\frac{\partial\phi_{e}}{\partial\xi}-\sigma_{i}\frac{\partial\phi_{i}}{\partial\xi}=0,\qquad{\rm at}\ \xi=\xi_{0}.\label{eq:reduced charge conservation}\end{eqnarray}
Indeed a consistency check \emph{a posteriori} justifies the simplification.
Far away from the droplet surface the electric field is uniform,\begin{equation}
-\nabla\phi_{e}=E_{0}\mathbf{\boldsymbol{z}},\qquad{\rm at}\ \xi\rightarrow\infty.\label{eq:farfield electric}\end{equation}
We also require that $\phi_{i}$ remains finite at $\xi=1$. For the
initial condition, we assume both the electric potential and the normal
component of the displacement vector are continuous:\begin{equation}
\epsilon_{e}\frac{\partial\phi_{e}}{\partial\xi}=\epsilon_{i}\frac{\partial\phi_{i}}{\partial\xi},\quad\phi_{e}=\phi_{i},\qquad{\rm at}\ \xi=\xi_{0},\; t=0.\end{equation}

Solutions for the electric potentials have been obtained previously
without including the displacement current \cite{Taylor1966,Bentenitis2005}.
With its inclusion the approach is similar and the results are \begin{equation}
\phi_{e}=E_{0}r_{0}\left[-\lambda\xi+\alpha Q_{1}(\xi)\right]\eta,\label{exterior potential}\end{equation}
\begin{equation}
\phi_{i}=E_{0}r_{0}\beta\xi\eta.\label{interior potential}\end{equation}
Here, $Q_{1}(\xi)$ is a 1st-degree Legendre polynomial of the second
kind. $\lambda\equiv c/r_{0}$ is the dimensionless semi-focal length.
The coefficients $\alpha$ and $\beta$ are determined by the interfacial
matching conditions (\ref{eq:potential continuous}) and (\ref{eq:reduced charge conservation})
which gives\begin{equation}
\alpha=\frac{\beta\xi_{0}+\lambda\xi_{0}}{Q_{1}(\xi_{0})},\label{eq:at}\end{equation}
\begin{widetext}
\begin{eqnarray}
\frac{\tau_{1}}{\tau_{2}}\left[\frac{Q_{1}^{'}(\xi_{0})\xi_{0}}{Q_{1}(\xi_{0})}-\frac{1}{\epsilon_{r}}\right]\frac{d\beta}{d\tau}+\left[\frac{\tau_{1}}{\tau_{2}}\left(\frac{Q_{1}^{'}(\xi_{0})}{Q_{1}(\xi_{0})}+\frac{Q_{1}^{''}(\xi_{0})Q_{1}(\xi_{0})-Q_{1}^{'2}(\xi_{0})}{Q_{1}^{2}(\xi_{0})}\xi_{0}\right)\frac{d\xi_{0}}{d\tau}-\frac{\tau_{1}}{\tau_{2}}\left(\frac{Q_{1}^{'}(\xi_{0})\xi_{0}}{Q_{1}(\xi_{0})\lambda}-\frac{1}{\epsilon_{r}\lambda}\right)\frac{d\lambda}{d\xi_{0}}\frac{d\xi_{0}}{d\tau}\right. \nonumber \\
\left.+\frac{Q_{1}^{'}(\xi_{0})\xi_{0}}{Q_{1}(\xi_{0})}-\frac{1}{\sigma_{r}}\right]\beta+\lambda\left[\frac{Q_{1}^{'}(\xi_{0})\xi_{0}}{Q_{1}(\xi_{0})}-1\right]+\frac{\tau_{1}}{\tau_{2}}\left[\frac{Q_{1}^{'}(\xi_{0})}{Q_{1}(\xi_{0})}+\frac{Q_{1}^{''}(\xi_{0})Q_{1}(\xi_{0})-Q_{1}^{'2}(\xi_{0})}{Q_{1}^{2}(\xi_{0})}\xi_{0}\right]\lambda\frac{d\xi_{0}}{d\tau}=0,\label{eq:bt}\end{eqnarray}
\end{widetext}
\begin{equation}
\alpha(0)=\lambda\xi_{0}\left(\epsilon_{r}-1\right),\quad\beta(0)=\frac{\epsilon_{r}\lambda\left(Q_{1}(\xi_{0})-Q_{1}^{'}(\xi_{0})\xi_{0}\right)}{\epsilon_{r}Q_{1}^{'}(\xi_{0})\xi_{0}-Q_{1}(\xi_{0})}.\end{equation}
Here $\epsilon_{r}\equiv\epsilon_{e}/\epsilon_{i}$ and $\sigma_{r}\equiv\sigma_{e}/\sigma_{i}$
are the permittivity ratio and the conductivity ratio, respectively.
$\tau_{1}\equiv\epsilon_{e}/\sigma_{e}$ is an electrical charging
time. $\tau_{2}\equiv r_{0}\mu_{e}/\gamma$ is a characteristic
flow timescale used below in the hydrodynamic problem, and $\gamma$ is the coefficient of surface tension. In the above
equations, a dimensionless time $\tau\equiv t/\tau_{2}$ has been
used. In general, Eq. (\ref{eq:bt}) needs to be integrated together
with an ODE for $\xi_{0}$ to obtain $\alpha$ and $\beta$. However,
in the limit of instantaneous-charge-relaxation time, $\tau_{1}/\tau_{2}\rightarrow0$,
and Eq. (\ref{eq:bt}) can be simplified to be\begin{equation}
\left[\frac{Q_{1}^{'}(\xi_{0})\xi_{0}}{Q_{1}(\xi_{0})}-\frac{1}{\sigma_{r}}\right]\beta+\lambda\left[\frac{Q_{1}^{'}(\xi_{0})\xi_{0}}{Q_{1}(\xi_{0})}-1\right]=0.\label{eq:simplified beta}\end{equation}
This result is equivalent to a solution employing the simplified boundary
condition $||\sigma\nabla\phi\cdot\mathbf{\boldsymbol{n}}||=0$ in
place of Eq. (\ref{eq:charge conservation}).

The normal and tangential electrostatic stresses are given by,\begin{equation}
S_{\xi\xi}=\frac{\epsilon}{2}\left(E_{\xi}^{2}-E_{\eta}^{2}\right),\qquad S_{\xi\eta}=\epsilon E_{\xi}E_{\eta},\label{N/T electric stress}\end{equation}
where $E_{\xi}=-(\partial\phi/\partial\xi)/h_{\xi}$ and $E_{\eta}=-(\partial\phi/\partial\eta)/h_{\eta}$
are the normal and tangential electric fields, respectively. $h_{\eta}$
is a metric coefficient of the prolate spheroidal coordinate system.
These stresses can be evaluated with the solutions (\ref{exterior potential})
and (\ref{interior potential}), and will be used in the stress matching
conditions below.

\subsection{The hydrodynamic problem}

In the regime of low-Reynolds-number flow, the governing equation
for the hydrodynamic problem can be rewritten in terms of the stream
function, $\psi$, as \begin{equation}
\rm{E}^{4}\psi=0.\label{stream function}\end{equation}
Here, the expression for the operator $\rm{E}^{2}$ can be found in Dubash and Mestel \cite{Dubash2007} and Bentenitis and Krause \cite{Bentenitis2005}. The stream function is related to the velocity components as\begin{equation}
u=-\frac{1}{h_{\xi}h_{\theta}}\frac{\partial\psi}{\partial\xi},\qquad v=\frac{1}{h_{\eta}h_{\theta}}\frac{\partial\psi}{\partial\eta}.\label{velocity field}\end{equation}
$h_{\theta}$ is a metric coefficient of the prolate spheroidal coordinate
system. At the interface, $u$ and $v$ represent the tangential and
normal velocities, respectively, and they are required to be continuous
\begin{equation}
u_{e}=u_{i},\qquad v_{e}=v_{i},\qquad{\rm at}\ \xi=\xi_{0}.\label{eq:no-slip condition}\end{equation}
In addition, we prescribe a kinematic condition relating the interfacial
displacement to the normal velocity, \begin{equation}
v(\xi=\xi_{0},\:\eta)=\frac{r_{0}\left(1-\xi_{0}^{-2}\right)^{-5/6}}{3\xi_{0}^{2}}\frac{\left(1-3\eta^{2}\right)}{\sqrt{\xi_{0}^{2}-\eta^{2}}}\frac{d\xi_{0}}{dt}.\label{eq:kinematic equation}\end{equation}

The total force on the interface resulting from the electrical stress,
the hydrodynamic stress, and the surface tension should be balanced
at every point. However, this constraint is impossible to satisfy
exactly within the framework of spheroidal deformation. Various authors
developed reduced stress-balance conditions instead \cite{Taylor1964,Sherwood1988,Bentenitis2005,Dubash2007}.
Here we follow the integrated formulae proposed by Sherwood \cite{Sherwood1988}
and Dubash and Mestel \cite{Dubash2007}\begin{equation}
\int u\cdot\left(T_{\xi\eta}^{e}-T_{\xi\eta}^{i}+S_{\xi\eta}^{e}-S_{\xi\eta}^{i}\right)ds=0,\label{averaged tangential stress balance}\end{equation}
\begin{equation}
\int v\cdot\left(T_{\xi\xi}^{e}-T_{\xi\xi}^{i}+S_{\xi\xi}^{e}-S_{\xi\xi}^{i}-\gamma\left(\frac{1}{R_{1}}+\frac{1}{R_{2}}\right)\right)ds=0.\label{averaged normal stress balance}\end{equation}
Equations (\ref{averaged tangential stress balance}) and (\ref{averaged normal stress balance})
represent a global balance of the tangential and normal stresses,
respectively derived from energy principles. Here $T$ denotes the
hydrodynamic stress, $R_{1}$ and $R_{2}$ are the two principal radii
of the curvature, and the integration is carried over the interface.

The general solution to (\ref{stream function}) was proposed by Dassios \emph{et al}. \cite{Dassios1994}
using the method of semi-separation:\begin{eqnarray}
\psi=g_{0}(\xi)&&G_{0}(\eta)+g_{1}(\xi)G_{1}(\eta)\nonumber \\
&&+\sum_{n=2}^{\infty}\left[g_{n}(\xi)G_{n}(\eta)+h_{n}(\xi)H_{n}(\eta)\right].\label{general solution for stream function}\end{eqnarray}
Here $G_{n}$ and $H_{n}$ are Gegenbauer functions of the first and
second kind, respectively. $g_{n}$ and $h_{n}$ are linear combinations
of $G_{n}$ and $H_{n}$. The detailed expressions for $G_{n}$, $H_{n}$,
$g_{n}$, and $h_{n}$ are found in Ref. \cite{Dassios1994}. Interested
readers are referred to Ref. \cite{Dassios1994} for further details.
After considering that the far field is quiescent, and that the velocities
remain finite at $\xi=1$, the stream functions can be simplified
to be \begin{eqnarray}
\psi_{e}=\sum_{n=1}^{\infty}\left[A_{2n+1}^{2n-1}H_{2n-1}(\xi)+A_{2n+1}^{2n+1}H_{2n+1}(\xi)\right.\nonumber \\
\left.+A_{2n+1}^{2n+3}H_{2n+3}(\xi)\right]G_{2n+1}(\eta),\label{exterior stream function}\end{eqnarray}
\begin{eqnarray}
\psi_{i}=\sum_{n=1}^{\infty}\left[B_{2n+1}^{2n-1}G_{2n-1}(\xi)+B_{2n+1}^{2n+1}G_{2n+1}(\xi)\right.\nonumber \\
\left.+B_{2n+1}^{2n+3}G_{2n+3}(\xi)\right]G_{2n+1}(\eta),\label{interior stream function}\end{eqnarray}
where $A$ and $B$ are unknown coefficients satisfying the relations
$A_{2n+1}^{2n+3}=A_{2n+3}^{2n+1}$, $B_{2n+1}^{2n+3}=B_{2n+3}^{2n+1}$.
In general, these coefficients are inter-dependent, and the full solution
can be obtained only with the entire infinite series. Here we seek
a truncated solution as an approximation, \begin{equation}
\psi_{e}=\left[A_{3}^{1}H_{1}(\xi)+A_{3}^{3}H_{3}(\xi)\right]G_{3}(\eta),\label{simplified exterior stream function}\end{equation}
\begin{equation}
\psi_{i}=\left[B_{3}^{3}G_{3}(\xi)+B_{3}^{5}G_{5}(\xi)\right]G_{3}(\eta).\label{simplified interior stream function}\end{equation}
Indeed, $G_{3}(\eta)$ gives a functional form in $\eta$ confirming
with that in Eq. (\ref{eq:kinematic equation}), which can be
rewritten as \begin{equation}
v(\xi=\xi_{0},\:\eta)=\frac{2c^{2}\sqrt{\xi_{0}-1}r_{0}\left(1-\xi_{0}^{-2}\right)^{-5/6}}{3\xi_{0}^{2}}\frac{G_{3}^{'}(\eta)}{h_{\eta}h_{\theta}}\frac{d\xi_{0}}{dt}.\label{G3}\end{equation}
This agreement in part validates the spheroidal shape assumption:
the shape represents the leading mode in the infinite series.

Equations (\ref{eq:no-slip condition}-\ref{averaged normal stress balance})
are combined to solve for the five unknown variables, namely, $A_{3}^{1}$,
$A_{3}^{3}$, $B_{3}^{3}$, $B_{3}^{5}$, and $\xi_{0}$. Specifically,
Eqs. (\ref{eq:no-slip condition}) and (\ref{eq:kinematic equation})
are first used to eliminate the $A_{3}^{1}$, $B_{3}^{3}$, $B_{3}^{5}$,\begin{equation}
A_{3}^{1}=H_{3}(\xi_{0})A_{3}^{3}-M\frac{d\xi_{0}}{dt},\label{A31}\end{equation}
\begin{equation}
B_{3}^{3}=\frac{-G_{5}(\xi_{0})H_{3}^{'}(\xi_{0})A_{3}^{3}+G_{5}^{'}(\xi_{0})M\frac{d\xi_{0}}{dt}}{N},\label{B33}\end{equation}
\begin{equation}
B_{3}^{5}=\frac{G_{3}(\xi_{0})H_{3}^{'}(\xi_{0})A_{3}^{3}-G_{3}^{'}(\xi_{0})M\frac{d\xi_{0}}{dt}}{N},\label{B35}\end{equation}
where $M\equiv2r_{0}^{3}/3(\xi_{0}^{3}-\xi_{0})$, and $N\equiv G_{3}(\xi_{0})G_{5}^{'}(\xi_{0})-G_{3}^{'}(\xi_{0})G_{5}(\xi_{0})$.
Further considering Eq. (\ref{averaged tangential stress balance}),
we can express $A_{3}^{3}$ in terms of $\xi_{0}$,
\begin{widetext}
\begin{equation}
A_{3}^{3}=\frac{cr_{0}^{2}\epsilon_{i}E_{0}^{2}\left\{ \xi_{0}\beta^{2}-\epsilon_{r}(\lambda-\alpha Q_{1}^{'}(\xi_{0}))(\lambda\xi_{0}-\alpha Q_{1}(\xi_{0}))\right\} f_{11}(\xi_{0})-\mu_{i}\left\{ (\mu_{r}-1)f_{12}(\xi_{0})+f_{13}(\xi_{0})\right\} M\frac{d\xi_{0}}{dt}}{-\mu_{i}\left\{ \mu_{r}f_{14}(\xi_{0})+f_{15}(\xi_{0})\right\} },\label{A33}\end{equation}
\end{widetext}
where $\mu_{r}\equiv\mu_{e}/\mu_{i}$ is the viscosity ratio. The
detailed expressions of $f_{11}(\xi_{0})-f_{15}(\xi_{0})$ are found
in the Appendix. This expression is inserted into Eq. (\ref{averaged normal stress balance})
to obtain the final result, an ODE governing the evolution of the
$\xi_{0}$,
\begin{widetext}
\addtocounter{equation}{0}\begin{subequations}\begin{equation}
\frac{d\xi_{0}}{d\tau}=-\frac{1}{F}\left[Q_{N}f_{21}(\xi_{0})+Q_{T}\frac{\mu_{r}f_{22}(\xi_{0})+f_{23}(\xi_{0})}{\mu_{r}f_{14}(\xi_{0})+f_{15}(\xi_{0})}-f_{24}(\xi_{0})\right],\label{drop shape evolution}\end{equation}
\begin{equation}
Q_{N}=\frac{Ca_{E}}{\lambda^{2}}\left[(\lambda-\alpha Q_{1}^{'}(\xi_{0}))^{2}+(\lambda-\alpha Q_{1}(\xi_{0})/\xi_{0})^{2}-2\beta^{2}/\epsilon_{r}\right],\label{eq:Qn}\end{equation}
\begin{equation}
Q_{T}=\frac{Ca_{E}}{\lambda^{2}}\left[(\lambda-\alpha Q_{1}^{'}(\xi_{0}))(\lambda-\alpha Q_{1}(\xi_{0})/\xi_{0})-\beta^{2}/\epsilon_{r}\right].\label{eq:Qt}\end{equation}
\end{subequations}
\end{widetext}
The detailed expressions of $f_{21}(\xi_{0})-f_{24}(\xi_{0})$,
and $F$ are also found in the Appendix. The coefficients $\alpha$
and $\beta$ are given by Eqs. (\ref{eq:at}) and (\ref{eq:bt}),
respectively. $Ca_{E}\equiv r_{0}\epsilon_{e}E_{0}^{2}/\gamma$ is
the electric capillary number. In Eq. (\ref{drop shape evolution}),
the three terms in the numerator on the RHS represent the contributions
from the normal stress, the tangential stress, and the surface tension,
respectively. At the equilibrium, the balance of the three forces
determines the final shape. The leading coefficients $Q_{N}$ and
$Q_{T}$ arise exclusively from the electrostatic stresses, and can
be used to estimate their respective influence on deformation. In
the limit of instantaneous relaxation, and by considering Eqs.
(\ref{eq:at}) and (\ref{eq:simplified beta}), $Q_{N}$ and $Q_{T}$
can be simplified to be \begin{equation}
Q_{N}=Ca_{E}K^{2}(\sigma_{r}^{2}+1-2\sigma_{r}^{2}/\epsilon_{r}),\\ Q_{T}=Ca_{E}K^{2}\sigma_{r}(1-\sigma_{r}/\epsilon_{r}),\label{eq:QtQn}\end{equation}
\begin{equation}
K\equiv\frac{Q_{1}(\xi_{0})-\xi_{0}Q_{1}^{'}(\xi_{0})}{Q_{1}(\xi_{0})-\sigma_{r}\xi_{0}Q_{1}^{'}(\xi_{0})}.\end{equation}
For this case, the evolution of $\xi_{0}$ is governed by a single
timescale, $\tau_{2}$. Once $\xi_{0}$ is obtained by solving the
Eqs. (\ref{eq:bt}) and (\ref{drop shape evolution}), the aspect
ratio is calculated by the formula\begin{equation}
\frac{a}{b}=(1-\xi_{0}^{-2})^{-\frac{1}{2}}.\label{eq:a/b}\end{equation}

\section{Comparison with previous results}

In this section, we compare our model prediction extensively with
results from previous work. The comparisons with theoretical/numerical
results and experimental data are respectively presented in Secs. \ref{sub:Comparison-with-previous-theory}
and \ref{sub:Comparison-with-experimental}.

\subsection{Comparison with previous theories and simulation\label{sub:Comparison-with-previous-theory}}

\begin{figure}
\center

\includegraphics[width=0.45\textwidth]{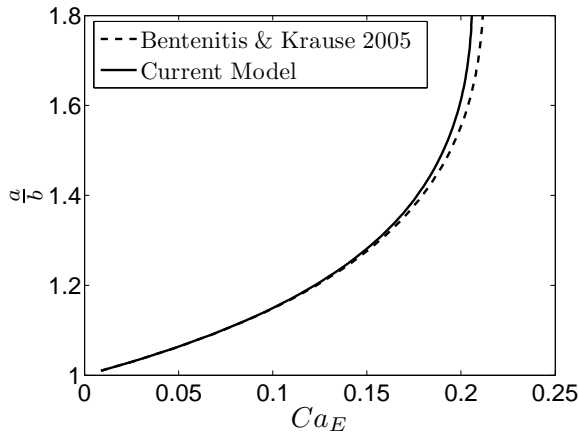}\caption{The equilibrium aspect ratio as a function of electric capillary number.
The parameters are $\sigma_{r}=1.19\times10^{-3}$, $\epsilon_{r}=3.24\times10^{-1}$,
and $\mu_{r}=7.33\times10^{-2}$.\label{fig:steady state aspect ratio}}

\end{figure}

We first consider the equilibrium shape, and compare our results with
those from Bentenitis and Krause \cite{Bentenitis2005}. For this case, the LHS of Eq.
(\ref{drop shape evolution}) is simply set to zero, resulting in
the so called discriminating equation,\begin{equation}
Q_{N}f_{21}(\xi_{0})+Q_{T}\frac{\mu_{r}f_{22}(\xi_{0})+f_{23}(\xi_{0})}{\mu_{r}f_{14}(\xi_{0})+f_{15}(\xi_{0})}=f_{24}(\xi_{0}).\label{steady state shape}\end{equation}
Here $Q_{N}$ and $Q_{T}$ are given by Eq. (\ref{eq:QtQn}).
$\xi_{0}$ is solved as a root(s) of this equation from which the
equilibrium aspect ratio, $a/b$, can be obtained. Equation (\ref{steady state shape})
shows that the equilibrium shape is only determined by the dimensionless
parameters $Ca_{E},\:\sigma_{r},\:\epsilon_{r},$ and $\mu_{r}$.
A comparison with the theoretical prediction by Bentenitis and Krause \cite{Bentenitis2005}
is shown in Fig. \ref{fig:steady state aspect ratio}. Note that
in this earlier work, the authors solved for the equilibrium shape
directly without obtaining the transient solution. A good agreement
is observed, although a different stress matching condition has been
used by Bentenitis and Krause \cite{Bentenitis2005} [see Eqs. (38) and (45) therein].

We next compare with the results from Dubash and Mestel \cite{Dubash2007}. In this
work, the authors developed a theoretical model, also with the spheroidal
shape assumption, to predict the transient deformation of a conducting,
inviscid droplet immersed in a viscous, nonconductive solution. This
special consideration leads to significant simplifications: both the
electric and hydrodynamic fields are absent within the droplet.
In addition, at the equilibrium state (if one is permitted), the hydrodynamic
flow outside the droplet is also quiescent, giving rise to the phenomenon
termed EHS. 

\begin{figure}
\center

\includegraphics[width=0.45\textwidth]{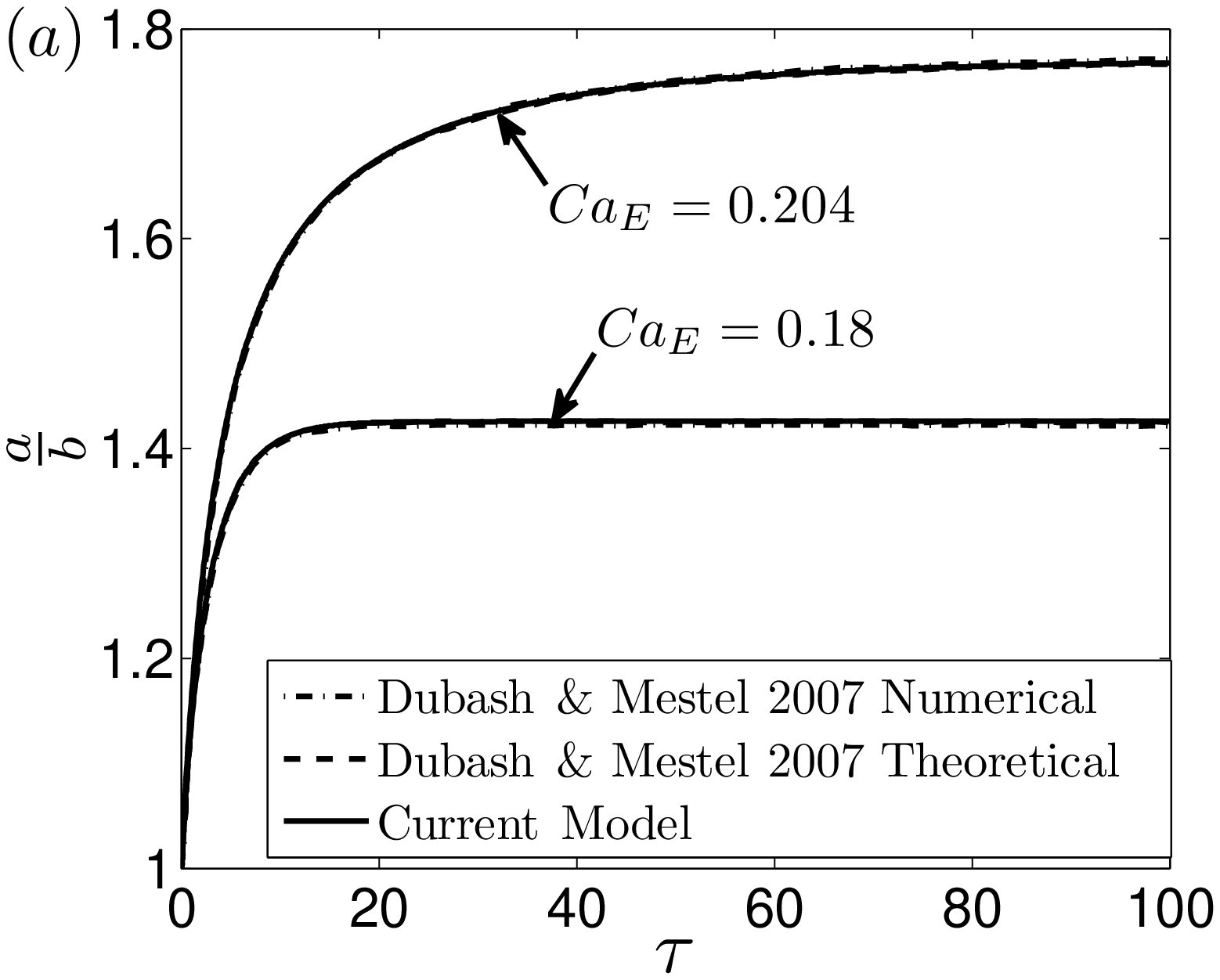}
\includegraphics[width=0.45\textwidth]{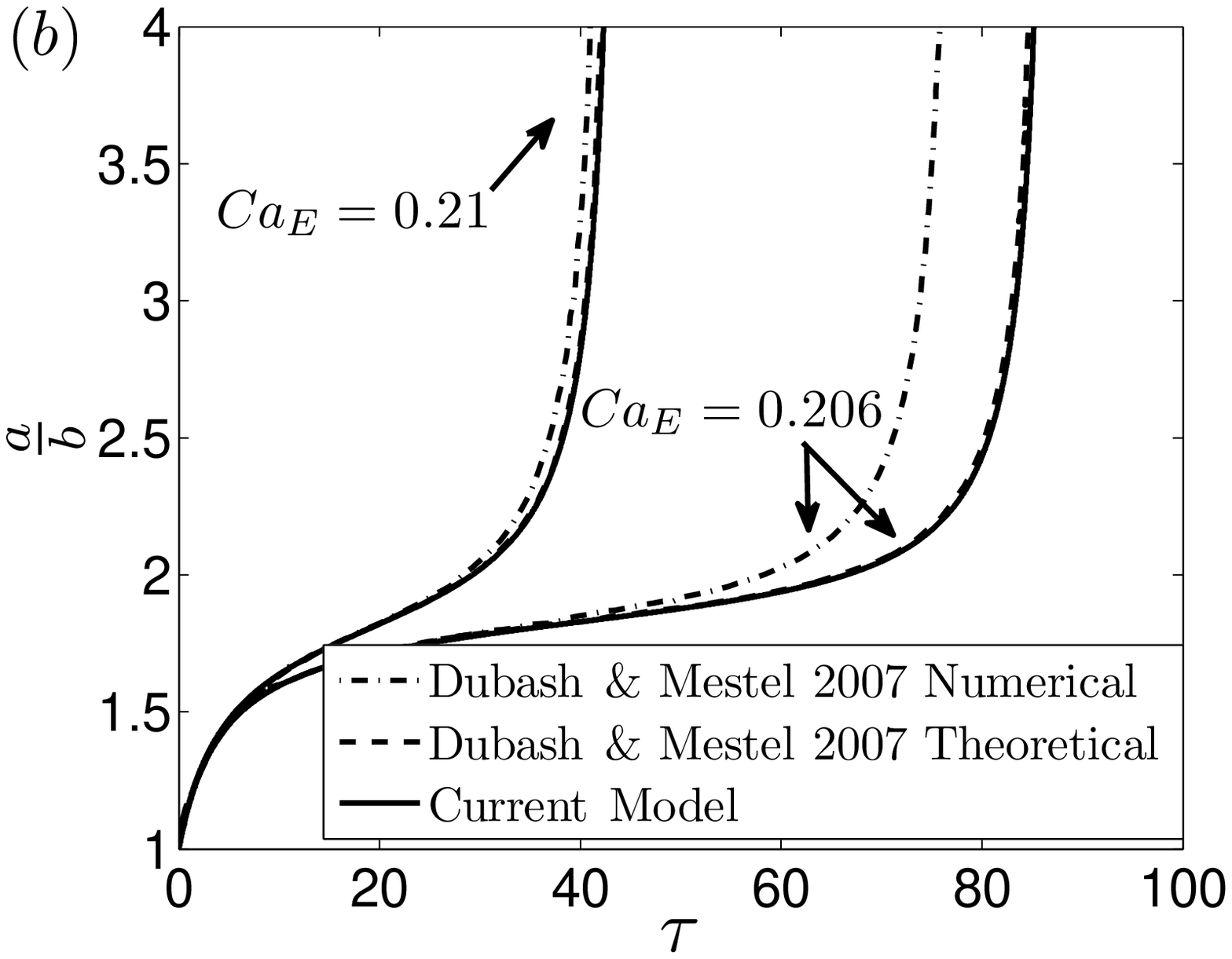}

\caption{The deformation of a conducting droplet in a highly viscous medium.
(a) $Ca_{E}=0.18$ and 0.204. (b) $Ca_{E}=0.206$ and 0.21. The dimensionless
time $\tau$ is defined as $\tau=t/\tau_{2}$, where $\tau_{2}=r_{0}\mu_{e}/\gamma$.\label{fig:evolution of infinite conducting drop}}

\end{figure}

In our generalized framework, the solution for this case is simply
achieved by setting $\sigma_{r}\rightarrow0$ and $\mu_{r}\rightarrow\infty$
in Eqs. (\ref{drop shape evolution}) and (\ref{eq:QtQn}). Note
that $\sigma_{r}\rightarrow0$ directly leads to instantaneous charge
relaxation. The resulting comparisons are shown in Fig. \ref{fig:evolution of infinite conducting drop}
in which the aspect ratio ($a/b$) is plotted as a function of time
for four different electric capillary numbers ($Ca_{E}$). For the
two lower values of $Ca_{E}$, the current model has excellent agreement
with both the theoretical and numerical predictions by Dubash and Mestel \cite{Dubash2007}
[Fig. \ref{fig:evolution of infinite conducting drop}(a)]. For these
$Ca_{E}$ values, final equilibria are achieved. As $Ca_{E}$ increases
[Fig. \ref{fig:evolution of infinite conducting drop}(b)], the deformation
becomes unstable and an equilibrium shape is no longer possible. The
rapid expansion with a sharp slope at the later stage preludes droplet
breakup. For these two cases, the theoretical models still agree with
each other, whereas some discrepancies exist with respect to the numerical
simulation, in particular for $Ca_{E}$=0.206. However, this discrepancy
is in general only noticeable when the $Ca_{E}$ number is above and
very close to the critical threshold of breakup ($Ca_{E}\sim$0.2044
for the case studied), due to a slight underprediction of the rate
of deformation by the theoretical models. A similar trend is observed
when comparing with the numerical simulation by Hirata \emph{et al}. \cite{Hirata2000}
(not shown). Overall, our model can serve as a good approximation
to the numerical model which is considered more accurate. 

\subsection{Comparison with experimental data\label{sub:Comparison-with-experimental}}

\begin{figure}
\center\includegraphics[width=0.45\textwidth]{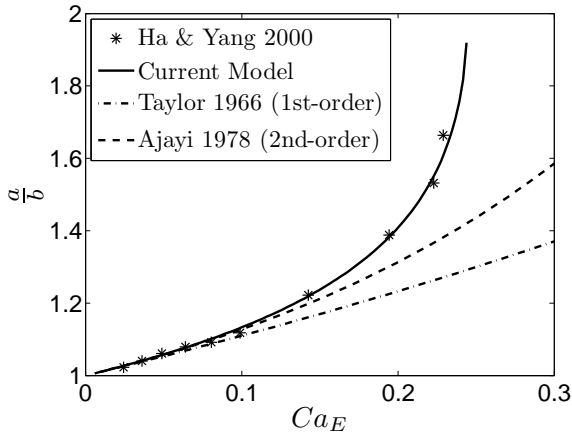}

\caption{The prediction from current model is compared with the small deformation
theories \cite{Taylor1966,Ajayi1978} and experimental
data \cite{Ha2000}. The parameters are $\sigma_{r}=0.03$, $\epsilon_{r}=0.73$,
and $\mu_{r}=1.14$.\label{fig:Drop-deformation-NN21}}

\end{figure}

The main source of experimental data comes from Ha and Yang \cite{Ha2000}. We
also begin with an examination of the final aspect ratio when an equilibrium
shape can be achieved. Figure \ref{fig:Drop-deformation-NN21} shows
the equilibrium aspect ratio of a castor oil droplet immersed in silicone
oil from Ha and Yang \cite{Ha2000}, as well as predicted by various models.
The current prediction is shown as a solid line, whereas the results
from first-order \cite{Taylor1966} and second-order \cite{Ajayi1978}
theories are shown as dot-dashed and dashed lines, respectively. Following
Lac and Homsy \cite{Lac2007}, we rescale $Ca_{E}$ to best match Ajayi's second-order
correction. This rescaling is equivalent to adjusting the surface
tension from $\gamma=3.3\times10^{-3}\;{\rm N}/{\rm m}$ used by Ha and Yang \cite{Ha2000}
(which is a fitting parameter in that work) to $\gamma=4.3\times10^{-3}\;{\rm N}/{\rm m}$.
The latter value is close to the lower bound, $\gamma=4.5\times10^{-3}\;{\rm N}/{\rm m}$,
measured by Salipante and Vlahovska \cite{Salipante2010}. In addition, we use $\sigma_{r}=0.03$
according to the measurements by Torza \emph{et al}. \cite{Torza1971}, Vizika and Saville \cite{Vizika1992},
and Salipante and Vlahovska \cite{Salipante2010}, which is slightly different from the value
of $\sigma_{r}=0.04$ used by Lac and Homsy \cite{Lac2007}. The results show good
agreement between the current model and the experimental data. Most
importantly, our theory correctly predicts a critical $Ca_{E}$ ($\sim$0.244)
for droplet breakup. In contrast, the small deformation theories can
not capture this critical phenomenon. 

We have also compared our theoretical prediction with the experimental
data from Bentenitis and Krause \cite{Bentenitis2005}, which measured the equilibrium
aspect ratio of a DGEBA droplet immersed in a PDMS solution. Since
our result is in good agreement with the theoretical prediction in
the same work (see Fig. \ref{fig:steady state aspect ratio}), which
in turn agrees well with the data, the comparison is not shown here
for brevity.%

\begin{figure}
\center

\includegraphics[width=0.45\textwidth]{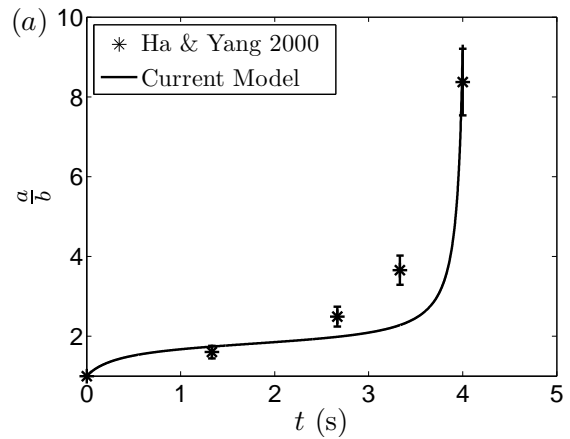}
\includegraphics[width=0.45\textwidth]{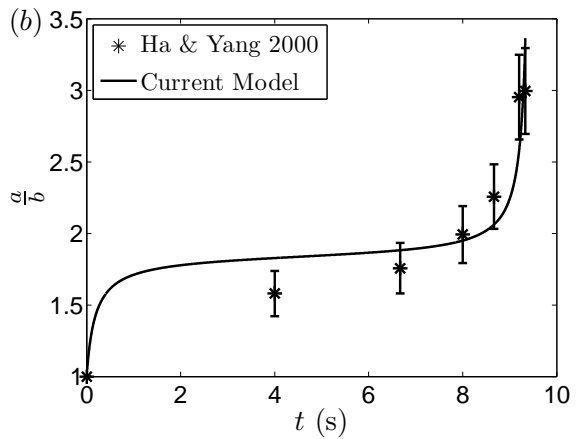}

\caption{Comparison of transient droplet deformation. (a) A water droplet in
silicone oil. The parameters are $\sigma_{r}=1\times10^{-6}$, $\epsilon_{r}=3.55\times10^{-2}$,
$\mu_{r}=1000$, $E_{0}=3.2\;{\rm kV/cm}$, $r_{0}=0.25\;{\rm cm}$,
$\mu_{e}=0.98\;{\rm Pa\cdot s}$, $\epsilon_{e}=2.478\times10^{-11}\;{\rm F}/{\rm m}$,
and $\gamma=3.037\times10^{-2}\;{\rm N}/{\rm m}$. (b) A water-ethanol
droplet in silicone oil. The parameters are $\sigma_{r}=1\times10^{-5}$,
$\epsilon_{r}=0.05$, $\mu_{r}=23.3$, $E_{0}=4.5\;{\rm kV/cm}$,
$r_{0}=0.14\;{\rm cm}$, $\mu_{e}=0.98\;{\rm Pa\cdot s}$, $\epsilon_{e}=2.478\times10^{-11}\;{\rm F}/{\rm m}$,
and $\gamma=3.432\times10^{-2}\;{\rm N}/{\rm m}$.\label{fig:evolution of conducting drop} }

\end{figure}

\begin{figure*}
\center

\includegraphics[width=0.7\textwidth]{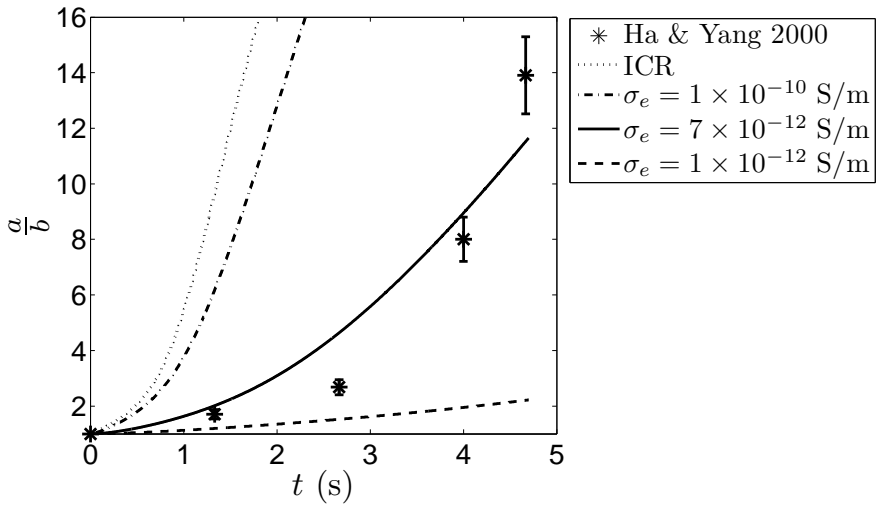}

\caption{Droplet deformation in the limit of extremely low conductivities.
The parameters are $\sigma_{r}=0.03$, $\epsilon_{r}=0.73$, $\mu_{r}=1.14$,
$E_{0}=3.2\;{\rm kV/cm}$, $r_{0}=0.16\;{\rm cm}$, $\mu_{e}=0.9\;{\rm Pa\cdot s}$,
$\epsilon_{e}=2.478\times10^{-11}\;{\rm F}/{\rm m}$, and $\gamma=5\times10^{-3}\;{\rm N}/{\rm m}$.
The best agreement between the data and the theory is found for $\sigma_{e}=7\times10^{-12}\;{\rm S/m}$.
For reference, the dotted line shows the calculation according to
the instantaneous-charge-relaxation (ICR) model.\label{fig:evolution of slightly conducting drop}}

\end{figure*}

Next, we will compare the transient solution from our model with data
from Ha and Yang \cite{Ha2000}. In Fig. \ref{fig:evolution of conducting drop}(a),
the data is extracted from Fig. 3 in the latter work, which captures
the deformation of a water droplet in silicone oil. The droplet is
fitted with an ellipse at every instant, based on which the aspect
ratio is calculated. A 10\% fitting error is estimated, and is shown
as error bars in Fig. \ref{fig:evolution of conducting drop}(a)
[the same approach is adopted to extract the data presented in Figs.
\ref{fig:evolution of conducting drop}(b) and \ref{fig:evolution of slightly conducting drop}].
The model prediction is calculated with Eqs. (\ref{drop shape evolution})
and (\ref{eq:QtQn}), and with $\sigma_{r}=1\times10^{-6}$, $\epsilon_{r}=3.55\times10^{-2}$,
$\mu_{r}=1000$, $E_{0}=3.2\;{\rm kV/cm}$, $r_{0}=0.25\;{\rm cm}$,
and $\mu_{e}=0.98\;{\rm Pa\cdot s}$ all directly taken from Ha and Yang \cite{Ha2000}.
For medium permittivity, we use $\epsilon_{e}=2.478\times10^{-11}\;{\rm F}/{\rm m}$
following the measurements by Torza \emph{et al}. \cite{Torza1971}, Vizika and Saville \cite{Vizika1992},
and Salipante and Vlahovska \cite{Salipante2010}. For surface tension, we use $\gamma=3.037\times10^{-2}\;{\rm N}/{\rm m}$,
which is consistent with the values reported by Torza \emph{et al}. \cite{Torza1971} and Vizika and Saville \cite{Vizika1992}.
In this case, the model is able to predict the deformation process
with good quantitative accuracy. In Fig. \ref{fig:evolution of conducting drop}(b),
a similar comparison is shown for a water-ethanol droplet in silicone
oil. The data is based on Fig. 4 in Ref. \cite{Ha2000}. For our calculation,
$\sigma_{r}=1\times10^{-5}$, $\epsilon_{r}=0.05$, $\mu_{r}=23.3$,
$E_{0}=4.5\;{\rm kV/cm}$, $r_{0}=0.14\;{\rm cm}$, $\mu_{e}=0.98\;{\rm Pa\cdot s}$,
and $\epsilon_{e}=2.478\times10^{-11}\;{\rm F}/{\rm m}$. Because
the droplet is doped with polyvinylpyrrolidone (a polymer solution),
the surface tension is not directly available, and is used as a fitting
parameter instead to generate the best agreement between theory and
data. The resulting value is $\gamma=3.432\times10^{-2}\;{\rm N}/{\rm m}$,
11\% higher than that for water/silicone oil which is used in Fig.
\ref{fig:evolution of conducting drop}(a). 

In contrast to the regime of instantaneous charge relaxation examined
in Fig. \ref{fig:evolution of conducting drop}, Fig. \ref{fig:evolution of slightly conducting drop}
represents droplet deformation in the finite-charging-time regime.
The data is extracted from Fig. 7 in Ref. \cite{Ha2000}. For this case,
the droplet is made of castor oil, and is immersed in silicone oil.
The extremely low conductivities of these media lead to a charging
time ($\sim$seconds) comparable to the deformation time, and the
full model, Eqs. (\ref{drop shape evolution})-(\ref{eq:Qt}),
has to be used. For our calculation, $\sigma_{r}=0.03$, $\epsilon_{r}=0.73$,
$\mu_{r}=1.14$, $E_{0}=3.2\;{\rm kV/cm}$, $r_{0}=0.16\;{\rm cm}$,
$\mu_{e}=0.9\;{\rm Pa\cdot s}$, $\epsilon_{e}=2.478\times10^{-11}\;{\rm F}/{\rm m}$,
and $\gamma=5\times10^{-3}\;{\rm N}/{\rm m}$. Note that the values
for the surface tension and the conductivity ratio follow the measurements
by Torza \emph{et al}. \cite{Torza1971}, Vizika and Saville \cite{Vizika1992},
and Salipante and Vlahovska \cite{Salipante2010}
which are believed to be more accurate than the original values of
$\sigma_{r}=0.1$ and $\gamma=3.3\times10^{-3}\;{\rm N}/{\rm m}$ given
by Ha and Yang \cite{Ha2000}. In addition, the actual conductivity of silicone
oil varies from $10^{-10}\;{\rm S/m}$ to $10^{-13}\;{\rm S/m}$ in
the literature \cite{Parkand2009,Salipante2010,Kim2011}.
In Fig. \ref{fig:evolution of slightly conducting drop}, we show
the calculation with three representative values within this range,
namely, $\sigma_{e}=1\times10^{-10}$, $7\times10^{-12}$, and $1\times10^{-12}\;{\rm S/m}$.
The best agreement is found for $\sigma_{e}=7\times10^{-12}\;{\rm S/m}$.
For comparison, the calculation according to the instantaneous-charge-relaxation
model [Eqs. (\ref{drop shape evolution}) and (\ref{eq:QtQn})]
is also shown, and is denoted by ICR. This simplified model clearly
overpredicts deformation by a significant degree.

In general, our model agrees well with experimental data in both steady
and transient states, and for a large parametric range. These comparisons
provide a strong validation for our model.

\section{The effects of stresses on deformation}

\begin{figure}
\center

\includegraphics[width=0.45\textwidth]{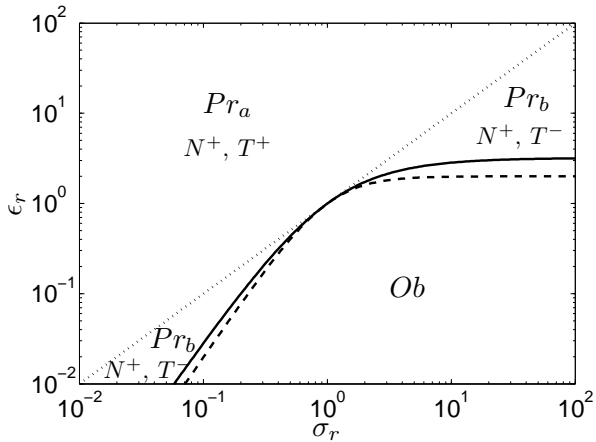}\caption{Phase diagram for droplet deformation. Here the dotted line is calculated
by satisfying $Q_{T}=0$. The solid line is calculated by solving for the root of Taylor's
discriminating function. The dashed line represents $Q_{N}=0$. $Pr$
and $Ob$ denote prolate and oblate deformation, respectively. $N$
and $T$ denote the effect of normal and tangential stresses, respectively,
and a $'+'$ or $'-'$ sign denotes facilitating or suppressing, respectively.\label{fig:sign-of-stress}}

\end{figure}

In this section, we demonstrate the utility of our theoretical results
by analyzing in-depth the governing equation. For simplicity, we focus
on the regime of instantaneous relaxation, where $Q_{N}$ and $Q_{T}$
are given by Eq. (\ref{eq:QtQn}). A main contribution of the
current work is that Eq. (\ref{drop shape evolution}) clearly
separates the effects by different forces. In the numerator of the
RHS, the three terms represent respectively the effects of the normal
stresses (both electrical and hydrodynamic), the tangential stresses
(both electrical and hydrodynamic), and the surface tension. Furthermore,
all the functions in this equation are positive ($f_{14}$, $f_{15}$,
$f_{21}-f_{24}$, $F$), such that the signs of $Q_{N}$ and $Q_{T}$
completely determine whether the normal and tangential stresses would
promote or suppress deformation. Due to the inverse relationship between
$\xi_{0}$ and the aspect ratio, $a/b$ [see Eq. (\ref{eq:a/b})],
a positive $Q_{N}$ or $Q_{T}$ indicates a positive contribution.
Evidently, surface tension always resists deformation. Because $Q_{N}$
and $Q_{T}$ depend exclusively on the electrical properties in a
simple manner [see Eq. (\ref{eq:QtQn})], their influences can
be conveniently analyzed using a phase diagram shown in Fig. \ref{fig:sign-of-stress}.
The dashed and dotted lines correspond to $Q_{N}=0$ and $Q_{T}=0$,
respectively. These lines separate the phase space into three regimes,
where $N$ and $T$ denote the normal and tangential stresses, and
the superscripts $'+'$ and $'-'$ denote a positive or negative contribution
to deformation, respectively. In addition, the solid line is obtained
by solving for the root of Taylor's discriminating function \cite{Taylor1966}, which
separates the prolate (denoted by $'Pr'$) and oblate (denoted
by $'Ob'$) regimes [this line can be equivalently obtained
by looking for the steady-state solution of $a/b=1$
from Eq. (\ref{steady state shape})]. 

Figure \ref{fig:sign-of-stress} can be used to shed light on the
physical processes governing deformation. First, the line for $Q_{T}=0$
separates the $T^{+}$ and $T^{-}$ regimes, which corroborates with
the previous results \cite{Taylor1966,Lac2007}. On this
dividing line, the velocity field becomes zero, so does the tangential
electrical stress. In Ref. \cite{Lac2007}, the viscosity ratio has opposite
effects on deformation in the $T^{+}$ and $T^{-}$ regimes. This
behavior is clearly explained by Eq. (\ref{drop shape evolution}).
Second, there is a small region within the oblate regime, namely,
the area between the solid and dashed lines where $Q_{N}$ is positive.
This suggests that the normal stress still tends to stretch the droplet
along the direction of the applied field. However, because $Q_{T}$
is negative, the tangential stresses overcome the normal stresses,
and stretch the droplet into an oblate shape. This new insight is
not available from previous analysis or simulations.

\begin{figure}
\center

\includegraphics[width=0.45\textwidth]{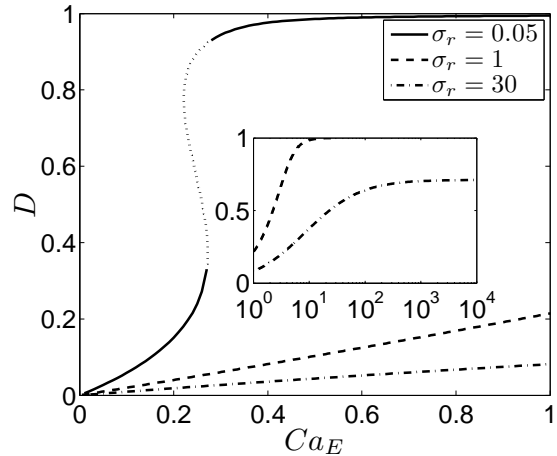}\caption{The behavior of equilibrium droplet deformation in different regimes.
For $\sigma_{r}=0.05$, $Q_{T}>0$; $\sigma_{r}=1$, $Q_{T}=0$; $\sigma_{r}=30$,
$Q_{T}<0$. As $Ca_{E}\rightarrow\infty$, an equilibrium shape is
only possible in the $T^{-}$ regime. Other parameters are $\epsilon_{r}=10$
and $\mu_{r}=1$.\label{fig:effect-of-conductivity}}

\end{figure}

Third, in the prolate regime where $Q_{N}$ is always positive, the
sign of $Q_{T}$ leads to different deformation behavior. Figure \ref{fig:effect-of-conductivity}
shows the equilibrium aspect ratio as a function of $Ca_{E}$ for
three specific cases. Note that the new variable\begin{equation}
D=\frac{a-b}{a+b}.\end{equation}
In this new definition, $D=0$ corresponds to $a/b=1$, and $D=1$
corresponds to $a/b\rightarrow\infty$. For all three cases, $\epsilon_{r}=10$
and $\mu_{r}=1$. For $\sigma_{r}=0.05$, $Q_{T}>0$. We observe hysteresis,
and $D$ approaches 1 rapidly in the upper brunch. The cases of $\sigma_{r}=1$
and $\sigma_{r}=30$ correspond to $Q_{T}=0$ and $Q_{T}<0$, respectively.
In general, as $Q_{T}$ decreases, the deformation becomes weaker for
comparable $Ca_{E}$ values. Most interestingly, for $\sigma_{r}=30$
($Q_{T}<0$), $D$ converges to a value less than 1 in the limit of
$Ca_{E}\rightarrow\infty$. This means that even for the very large
applied electric field strength, a finite equilibrium aspect ratio
can be achieved. We emphasize this scenario is only possible in the
$T^{-}$ regime. For large $E_{0}$ values, corresponding to large
$Ca_{E}$, the resistive effect from surface tension is negligible,
and the only way to obtain a finite equilibrium aspect ratio is therefore
by balancing the normal and tangential stresses. Since $Q_{N}$ is
positive, $Q_{T}$ has to be negative.

\section{Conclusions}

In conclusion, we have developed a transient analysis to quantify
droplet deformation under DC electric fields. The full Taylor-Melcher
leaky dielectric model is employed where the charge relaxation time
is considered finite. In this framework, instantaneous charge relaxation
is treated as a special limiting case. The droplet is assumed to be
spheroidal in shape for all times. The main result is an ODE governing
the evolution of the droplet aspect ratio. The model is validated
by extensively comparing predicted deformation with both previous
theoretical and numerical studies, and with experimental data. In
particular, the experimental results by Ha and Yang \cite{Ha2000}, which were
obtained with extremely low medium conductivities are well captured
by the simulation with the finite-time charge-relaxation model. The
model is used to analyze the effects of parameters and stresses on
the deformation characteristics. The results demonstrate clearly that
in different regimes according to the sign of $Q_{T}$, the stresses
contribute qualitatively differently to deformation. Last but not
least, this work lays the foundation for the study of a more complex
problem, namely, vesicle electrodeformation and relaxation. This problem is the pursuit of our future work.

\begin{acknowledgments}
JZ and HL acknowledge fund support from an NSF award CBET-0747886 with Dr William Schultz
and Dr Henning Winter as contract monitors.
\end{acknowledgments}

\appendix

\section{Appendixes}

The functions $f_{11}(\xi_{0})-f_{15}(\xi_{0})$ in Eq. (\ref{A33})
are given in the following expressions:

\begin{equation}
f_{11}(\xi_{0})=\int\frac{G_{3}(\eta)\eta}{(\xi_{0}^{2}-\eta^{2})}d\eta,\end{equation}
\begin{equation}
f_{12}(\xi_{0})=\frac{1}{\xi_{0}^{2}-1}\left\{ \int\frac{G_{3}(\eta)\eta}{(\xi_{0}^{2}-\eta^{2})}\left(\frac{(1-3\eta^{2})}{(\xi_{0}^{2}-\eta^{2})}-3\right)d\eta\right\} ,\end{equation}
\begin{equation}
f_{13}(\xi_{0})=\frac{G_{3}^{''}(\xi_{0})G_{5}^{'}(\xi_{0})-G_{3}^{'}(\xi_{0})G_{5}^{''}(\xi_{0})}{2N}\cdot f_{11}(\xi_{0}),\end{equation}
\begin{equation}
f_{14}(\xi_{0})=-\xi_{0}H_{3}^{'}(\xi_{0})\int\frac{G_{3}(\eta)\eta}{(\xi_{0}^{2}-\eta^{2})^{2}}d\eta+\frac{1}{2}H_{3}^{''}(\xi_{0})f_{11}(\xi_{0}),\end{equation}
\begin{widetext}
\begin{equation}
f_{15}(\xi_{0})=-\frac{H_{3}^{'}(\xi_{0})\left[G_{3}(\xi_{0})G_{5}^{''}(\xi_{0})-G_{3}^{''}(\xi_{0})G_{5}(\xi_{0})\right]}{2N}f_{11}(\xi_{0})+\xi_{0}H_{3}^{'}(\xi_{0})\int\frac{G_{3}(\eta)\eta}{(\xi_{0}^{2}-\eta^{2})^{2}}d\eta.\end{equation}

The functions $f_{21}(\xi_{0})-f_{24}(\xi_{0})$ and $F$ in Eq.
(\ref{drop shape evolution}) are given in the following expressions:

\begin{equation}
f_{21}(\xi_{0})=\frac{1}{2}\xi_{0}^{2}\int\frac{(\eta^{2}-1)(3\eta^{2}-1)}{(\xi_{0}^{2}-\eta^{2})}d\eta,\end{equation}

\begin{equation}
f_{22}(\xi_{0})=\xi_{0}f_{11}(\xi_{0})\left[-H_{3}^{'}(\xi_{0})\int\frac{(1-3\eta^{2})(\xi_{0}^{2}-3\xi_{0}^{2}\eta^{2}+2\eta^{4})}{(\xi_{0}^{2}-\eta^{2})^{2}}d\eta+3\xi_{0}H_{3}(\xi_{0})\int\frac{1-3\eta^{2}}{(\xi_{0}^{2}-\eta^{2})}d\eta\right],\end{equation}

\begin{equation}
f_{23}(\xi_{0})=\xi_{0}f_{11}(\xi_{0})\left[-\frac{49(1-3\xi_{0}^{2})G_{3}(\xi_{0})H_{3}^{'}(\xi_{0})}{30N}+H_{3}^{'}(\xi_{0})\int\frac{(1-3\eta^{2})(\xi_{0}^{2}-3\xi_{0}^{2}\eta^{2}+2\eta^{4})}{(\xi_{0}^{2}-\eta^{2})^{2}}d\eta\right],\end{equation}
\begin{eqnarray}
f_{24}(\xi_{0})=\xi_{0}^{3}(1-\xi_{0}^{-2})^{\frac{5}{6}}\int\frac{3\eta^{2}-1}{(\xi_{0}^{2}-\eta^{2})^{\frac{3}{2}}}d\eta+\xi_{0}(1-\xi_{0}^{-2})^{-\frac{1}{6}}\int\frac{3\eta^{2}-1}{\sqrt{\xi_{0}^{2}-\eta^{2}}}d\eta,\end{eqnarray}
\begin{equation}
F=-\frac{2}{3}\left(f_{25}(\xi_{0})+f_{26}(\xi_{0})/\mu_{r}\right),\end{equation}
where
\begin{equation}
f_{25}(\xi_{0})=-\frac{f_{22}(\xi_{0})}{\xi_{0}f_{11}(\xi_{0})}\frac{(\mu_{r}-1)f_{12}(\xi_{0})+f_{13}(\xi_{0})}{\mu_{r}f_{14}(\xi_{0})+f_{15}(\xi_{0})}-3\xi_{0}\int\frac{3\eta^{2}-1}{(\xi_{0}^{2}-\eta^{2})}d\eta-\frac{\xi_{0}}{\xi_{0}^{2}-1}\int\frac{(2\xi_{0}^{2}-\eta^{2}-1)(1-3\eta^{2})^{2}}{(\xi_{0}^{2}-\eta^{2})^{2}}d\eta,\end{equation}
\begin{equation}
f_{26}(\xi_{0})=-\frac{f_{23}(\xi_{0})}{\xi_{0}f_{11}(\xi_{0})}\frac{(\mu_{r}-1)f_{12}(\xi_{0})+f_{13}(\xi_{0})}{\mu_{r}f_{14}(\xi_{0})+f_{15}(\xi_{0})}-\frac{49(1-3\xi_{0}^{2})G_{3}^{'}(\xi_{0})}{30N}+\frac{\xi_{0}}{\xi_{0}^{2}-1}\int\frac{(2\xi_{0}^{2}-\eta^{2}-1)(1-3\eta^{2})^{2}}{(\xi_{0}^{2}-\eta^{2})^{2}}d\eta.\end{equation}
\end{widetext}

\bibliography{reference_on_Droplet_deformation}

\begin{thebibliography}{27}%
\makeatletter
\providecommand \@ifxundefined [1]{%
 \@ifx{#1\undefined}
}%
\providecommand \@ifnum [1]{%
 \ifnum #1\expandafter \@firstoftwo
 \else \expandafter \@secondoftwo
 \fi
}%
\providecommand \@ifx [1]{%
 \ifx #1\expandafter \@firstoftwo
 \else \expandafter \@secondoftwo
 \fi
}%
\providecommand \natexlab [1]{#1}%
\providecommand \enquote  [1]{``#1''}%
\providecommand \bibnamefont  [1]{#1}%
\providecommand \bibfnamefont [1]{#1}%
\providecommand \citenamefont [1]{#1}%
\providecommand \href@noop [0]{\@secondoftwo}%
\providecommand \href [0]{\begingroup \@sanitize@url \@href}%
\providecommand \@href[1]{\@@startlink{#1}\@@href}%
\providecommand \@@href[1]{\endgroup#1\@@endlink}%
\providecommand \@sanitize@url [0]{\catcode `\\12\catcode `\$12\catcode
  `\&12\catcode `\#12\catcode `\^12\catcode `\_12\catcode `\%12\relax}%
\providecommand \@@startlink[1]{}%
\providecommand \@@endlink[0]{}%
\providecommand \url  [0]{\begingroup\@sanitize@url \@url }%
\providecommand \@url [1]{\endgroup\@href {#1}{\urlprefix }}%
\providecommand \urlprefix  [0]{URL }%
\providecommand \Eprint [0]{\href }%
\providecommand \doibase [0]{http://dx.doi.org/}%
\providecommand \selectlanguage [0]{\@gobble}%
\providecommand \bibinfo  [0]{\@secondoftwo}%
\providecommand \bibfield  [0]{\@secondoftwo}%
\providecommand \translation [1]{[#1]}%
\providecommand \BibitemOpen [0]{}%
\providecommand \bibitemStop [0]{}%
\providecommand \bibitemNoStop [0]{.\EOS\space}%
\providecommand \EOS [0]{\spacefactor3000\relax}%
\providecommand \BibitemShut  [1]{\csname bibitem#1\endcsname}%
\let\auto@bib@innerbib\@empty
\bibitem [{\citenamefont {Wu}\ and\ \citenamefont {Clark}(2008)}]{Wu2008}%
  \BibitemOpen
  \bibfield  {author} {\bibinfo {author} {\bibfnamefont {Y.}~\bibnamefont
  {Wu}}\ and\ \bibinfo {author} {\bibfnamefont {R.~L.}\ \bibnamefont {Clark}},\
  }\href@noop {} {\bibfield  {journal} {\bibinfo  {journal} {J. Biomater. Sci.
  Polymer Edn}\ }\textbf {\bibinfo {volume} {19}},\ \bibinfo {pages} {573}
  (\bibinfo {year} {2008})}\BibitemShut {NoStop}%
\bibitem [{\citenamefont {Kanazawa}\ \emph {et~al.}(2008)\citenamefont
  {Kanazawa}, \citenamefont {Takahashi},\ and\ \citenamefont
  {Nomoto}}]{Kanazawa2008}%
  \BibitemOpen
  \bibfield  {author} {\bibinfo {author} {\bibfnamefont {S.}~\bibnamefont
  {Kanazawa}}, \bibinfo {author} {\bibfnamefont {Y.}~\bibnamefont {Takahashi}},
  \ and\ \bibinfo {author} {\bibfnamefont {Y.}~\bibnamefont {Nomoto}},\
  }\href@noop {} {\bibfield  {journal} {\bibinfo  {journal} {IEEE Trans. Ind.
  Appl.}\ }\textbf {\bibinfo {volume} {44}},\ \bibinfo {pages} {1084} (\bibinfo
  {year} {2008})}\BibitemShut {NoStop}%
\bibitem [{\citenamefont {Basaran}(2002)}]{Basaran2002}%
  \BibitemOpen
  \bibfield  {author} {\bibinfo {author} {\bibfnamefont {O.~A.}\ \bibnamefont
  {Basaran}},\ }\href@noop {} {\bibfield  {journal} {\bibinfo  {journal} {AIChE
  J.}\ }\textbf {\bibinfo {volume} {48}},\ \bibinfo {pages} {1842} (\bibinfo
  {year} {2002})}\BibitemShut {NoStop}%
\bibitem [{\citenamefont {Allan}\ and\ \citenamefont
  {Mason}(1962)}]{Allan1962}%
  \BibitemOpen
  \bibfield  {author} {\bibinfo {author} {\bibfnamefont {R.~S.}\ \bibnamefont
  {Allan}}\ and\ \bibinfo {author} {\bibfnamefont {S.~G.}\ \bibnamefont
  {Mason}},\ }\href@noop {} {\bibfield  {journal} {\bibinfo  {journal} {Proc.
  R. Soc. Lond. A}\ }\textbf {\bibinfo {volume} {267}},\ \bibinfo {pages} {45}
  (\bibinfo {year} {1962})}\BibitemShut {NoStop}%
\bibitem [{\citenamefont {Taylor}(1964)}]{Taylor1964}%
  \BibitemOpen
  \bibfield  {author} {\bibinfo {author} {\bibfnamefont {G.~I.}\ \bibnamefont
  {Taylor}},\ }\href@noop {} {\bibfield  {journal} {\bibinfo  {journal} {Proc.
  R. Soc. Lond. A}\ }\textbf {\bibinfo {volume} {280}},\ \bibinfo {pages} {383}
  (\bibinfo {year} {1964})}\BibitemShut {NoStop}%
\bibitem [{\citenamefont {Miksis}(1981)}]{Miksis1981}%
  \BibitemOpen
  \bibfield  {author} {\bibinfo {author} {\bibfnamefont {M.~J.}\ \bibnamefont
  {Miksis}},\ }\href@noop {} {\bibfield  {journal} {\bibinfo  {journal} {Phys.
  Fluids}\ }\textbf {\bibinfo {volume} {24}},\ \bibinfo {pages} {1967}
  (\bibinfo {year} {1981})}\BibitemShut {NoStop}%
\bibitem [{\citenamefont {Sherwood}(1988)}]{Sherwood1988}%
  \BibitemOpen
  \bibfield  {author} {\bibinfo {author} {\bibfnamefont {J.~D.}\ \bibnamefont
  {Sherwood}},\ }\href@noop {} {\bibfield  {journal} {\bibinfo  {journal} {J.
  Fluid Mech.}\ }\textbf {\bibinfo {volume} {188}},\ \bibinfo {pages} {133}
  (\bibinfo {year} {1988})}\BibitemShut {NoStop}%
\bibitem [{\citenamefont {Ha}\ and\ \citenamefont {Yang}(2000)}]{Ha2000}%
  \BibitemOpen
  \bibfield  {author} {\bibinfo {author} {\bibfnamefont {J.-W.}\ \bibnamefont
  {Ha}}\ and\ \bibinfo {author} {\bibfnamefont {S.-M.}\ \bibnamefont {Yang}},\
  }\href@noop {} {\bibfield  {journal} {\bibinfo  {journal} {J. Fluid Mech.}\
  }\textbf {\bibinfo {volume} {405}},\ \bibinfo {pages} {131} (\bibinfo {year}
  {2000})}\BibitemShut {NoStop}%
\bibitem [{\citenamefont {Dubash}\ and\ \citenamefont
  {Mestel}(2007)}]{Dubash2007}%
  \BibitemOpen
  \bibfield  {author} {\bibinfo {author} {\bibfnamefont {N.}~\bibnamefont
  {Dubash}}\ and\ \bibinfo {author} {\bibfnamefont {A.~J.}\ \bibnamefont
  {Mestel}},\ }\href@noop {} {\bibfield  {journal} {\bibinfo  {journal} {J.
  Fluid Mech.}\ }\textbf {\bibinfo {volume} {581}},\ \bibinfo {pages} {469}
  (\bibinfo {year} {2007})}\BibitemShut {NoStop}%
\bibitem [{\citenamefont {Taylor}(1966)}]{Taylor1966}%
  \BibitemOpen
  \bibfield  {author} {\bibinfo {author} {\bibfnamefont {G.~I.}\ \bibnamefont
  {Taylor}},\ }\href@noop {} {\bibfield  {journal} {\bibinfo  {journal} {Proc.
  R. Soc. Lond. A}\ }\textbf {\bibinfo {volume} {291}},\ \bibinfo {pages} {159}
  (\bibinfo {year} {1966})}\BibitemShut {NoStop}%
\bibitem [{\citenamefont {Torza}\ \emph {et~al.}(1971)\citenamefont {Torza},
  \citenamefont {Cox},\ and\ \citenamefont {Mason}}]{Torza1971}%
  \BibitemOpen
  \bibfield  {author} {\bibinfo {author} {\bibfnamefont {S.}~\bibnamefont
  {Torza}}, \bibinfo {author} {\bibfnamefont {R.~G.}\ \bibnamefont {Cox}}, \
  and\ \bibinfo {author} {\bibfnamefont {S.~G.}\ \bibnamefont {Mason}},\
  }\href@noop {} {\bibfield  {journal} {\bibinfo  {journal} {Phil. Trans. R.
  Soc. Lond. A}\ }\textbf {\bibinfo {volume} {269}},\ \bibinfo {pages} {295}
  (\bibinfo {year} {1971})}\BibitemShut {NoStop}%
\bibitem [{\citenamefont {Ajayi}(1978)}]{Ajayi1978}%
  \BibitemOpen
  \bibfield  {author} {\bibinfo {author} {\bibfnamefont {O.~O.}\ \bibnamefont
  {Ajayi}},\ }\href@noop {} {\bibfield  {journal} {\bibinfo  {journal} {Proc.
  R. Soc. Lond. A}\ }\textbf {\bibinfo {volume} {364}},\ \bibinfo {pages} {499}
  (\bibinfo {year} {1978})}\BibitemShut {NoStop}%
\bibitem [{\citenamefont {Vizika}\ and\ \citenamefont
  {Saville}(1992)}]{Vizika1992}%
  \BibitemOpen
  \bibfield  {author} {\bibinfo {author} {\bibfnamefont {O.}~\bibnamefont
  {Vizika}}\ and\ \bibinfo {author} {\bibfnamefont {D.~A.}\ \bibnamefont
  {Saville}},\ }\href@noop {} {\bibfield  {journal} {\bibinfo  {journal} {J.
  Fluid Mech.}\ }\textbf {\bibinfo {volume} {239}},\ \bibinfo {pages} {1}
  (\bibinfo {year} {1992})}\BibitemShut {NoStop}%
\bibitem [{\citenamefont {Feng}\ and\ \citenamefont {Scott}(1996)}]{Feng1996}%
  \BibitemOpen
  \bibfield  {author} {\bibinfo {author} {\bibfnamefont {J.~Q.}\ \bibnamefont
  {Feng}}\ and\ \bibinfo {author} {\bibfnamefont {T.~C.}\ \bibnamefont
  {Scott}},\ }\href@noop {} {\bibfield  {journal} {\bibinfo  {journal} {J.
  Fluid Mech.}\ }\textbf {\bibinfo {volume} {311}},\ \bibinfo {pages} {289}
  (\bibinfo {year} {1996})}\BibitemShut {NoStop}%
\bibitem [{\citenamefont {Baygents}\ \emph {et~al.}(1998)\citenamefont
  {Baygents}, \citenamefont {Rivette},\ and\ \citenamefont
  {Stone}}]{Baygents1998}%
  \BibitemOpen
  \bibfield  {author} {\bibinfo {author} {\bibfnamefont {J.~C.}\ \bibnamefont
  {Baygents}}, \bibinfo {author} {\bibfnamefont {N.~J.}\ \bibnamefont
  {Rivette}}, \ and\ \bibinfo {author} {\bibfnamefont {H.~A.}\ \bibnamefont
  {Stone}},\ }\href@noop {} {\bibfield  {journal} {\bibinfo  {journal} {J.
  Fluid Mech.}\ }\textbf {\bibinfo {volume} {368}},\ \bibinfo {pages} {359}
  (\bibinfo {year} {1998})}\BibitemShut {NoStop}%
\bibitem [{\citenamefont {Feng}(1999)}]{Feng1999}%
  \BibitemOpen
  \bibfield  {author} {\bibinfo {author} {\bibfnamefont {J.~Q.}\ \bibnamefont
  {Feng}},\ }\href@noop {} {\bibfield  {journal} {\bibinfo  {journal} {Proc. R.
  Soc. Lond. A}\ }\textbf {\bibinfo {volume} {455}},\ \bibinfo {pages} {2245}
  (\bibinfo {year} {1999})}\BibitemShut {NoStop}%
\bibitem [{\citenamefont {Hirata}\ \emph {et~al.}(2000)\citenamefont {Hirata},
  \citenamefont {Kikuchi}, \citenamefont {Tsukada},\ and\ \citenamefont
  {Hozawa}}]{Hirata2000}%
  \BibitemOpen
  \bibfield  {author} {\bibinfo {author} {\bibfnamefont {T.}~\bibnamefont
  {Hirata}}, \bibinfo {author} {\bibfnamefont {T.}~\bibnamefont {Kikuchi}},
  \bibinfo {author} {\bibfnamefont {T.}~\bibnamefont {Tsukada}}, \ and\
  \bibinfo {author} {\bibfnamefont {M.}~\bibnamefont {Hozawa}},\ }\href@noop {}
  {\bibfield  {journal} {\bibinfo  {journal} {J. Chem. Engng Japan}\ }\textbf
  {\bibinfo {volume} {33}},\ \bibinfo {pages} {160} (\bibinfo {year}
  {2000})}\BibitemShut {NoStop}%
\bibitem [{\citenamefont {Bentenitis}\ and\ \citenamefont
  {Krause}(2005)}]{Bentenitis2005}%
  \BibitemOpen
  \bibfield  {author} {\bibinfo {author} {\bibfnamefont {N.}~\bibnamefont
  {Bentenitis}}\ and\ \bibinfo {author} {\bibfnamefont {S.}~\bibnamefont
  {Krause}},\ }\href@noop {} {\bibfield  {journal} {\bibinfo  {journal}
  {Langmuir}\ }\textbf {\bibinfo {volume} {21}},\ \bibinfo {pages} {6194}
  (\bibinfo {year} {2005})}\BibitemShut {NoStop}%
\bibitem [{\citenamefont {Sato}\ \emph {et~al.}(2006)\citenamefont {Sato},
  \citenamefont {Kaji}, \citenamefont {Mochizuki},\ and\ \citenamefont
  {Mori}}]{Sato2006}%
  \BibitemOpen
  \bibfield  {author} {\bibinfo {author} {\bibfnamefont {H.}~\bibnamefont
  {Sato}}, \bibinfo {author} {\bibfnamefont {N.}~\bibnamefont {Kaji}}, \bibinfo
  {author} {\bibfnamefont {T.}~\bibnamefont {Mochizuki}}, \ and\ \bibinfo
  {author} {\bibfnamefont {Y.~H.}\ \bibnamefont {Mori}},\ }\href@noop {}
  {\bibfield  {journal} {\bibinfo  {journal} {Phys. Fluids}\ }\textbf {\bibinfo
  {volume} {18}},\ \bibinfo {pages} {127101} (\bibinfo {year}
  {2006})}\BibitemShut {NoStop}%
\bibitem [{\citenamefont {Lac}\ and\ \citenamefont {Homsy}(2007)}]{Lac2007}%
  \BibitemOpen
  \bibfield  {author} {\bibinfo {author} {\bibfnamefont {E.}~\bibnamefont
  {Lac}}\ and\ \bibinfo {author} {\bibfnamefont {G.~M.}\ \bibnamefont
  {Homsy}},\ }\href@noop {} {\bibfield  {journal} {\bibinfo  {journal} {J.
  Fluid Mech.}\ }\textbf {\bibinfo {volume} {590}},\ \bibinfo {pages} {239}
  (\bibinfo {year} {2007})}\BibitemShut {NoStop}%
\bibitem [{\citenamefont {Melcher}\ and\ \citenamefont
  {Taylor}(1969)}]{Melcher1969}%
  \BibitemOpen
  \bibfield  {author} {\bibinfo {author} {\bibfnamefont {J.~R.}\ \bibnamefont
  {Melcher}}\ and\ \bibinfo {author} {\bibfnamefont {G.~I.}\ \bibnamefont
  {Taylor}},\ }\href@noop {} {\bibfield  {journal} {\bibinfo  {journal} {Annu.
  Rev. Fluid Mech.}\ }\textbf {\bibinfo {volume} {1}},\ \bibinfo {pages} {111}
  (\bibinfo {year} {1969})}\BibitemShut {NoStop}%
\bibitem [{\citenamefont {Saville}(1997)}]{Saville1997}%
  \BibitemOpen
  \bibfield  {author} {\bibinfo {author} {\bibfnamefont {D.~A.}\ \bibnamefont
  {Saville}},\ }\href@noop {} {\bibfield  {journal} {\bibinfo  {journal} {Annu.
  Rev. Fluid Mech.}\ }\textbf {\bibinfo {volume} {29}},\ \bibinfo {pages} {27}
  (\bibinfo {year} {1997})}\BibitemShut {NoStop}%
\bibitem [{\citenamefont {Zhang}\ \emph {et~al.}(2011)\citenamefont {Zhang},
  \citenamefont {Zahn},\ and\ \citenamefont {Lin}}]{Zhang2011}%
  \BibitemOpen
  \bibfield  {author} {\bibinfo {author} {\bibfnamefont {J.}~\bibnamefont
  {Zhang}}, \bibinfo {author} {\bibfnamefont {J.~D.}\ \bibnamefont {Zahn}}, \
  and\ \bibinfo {author} {\bibfnamefont {H.}~\bibnamefont {Lin}},\ }\href@noop
  {} {\bibfield  {journal} {\bibinfo  {journal} {J. Fluid Mech.}\ }\textbf
  {\bibinfo {volume} {681}},\ \bibinfo {pages} {293} (\bibinfo {year}
  {2011})}\BibitemShut {NoStop}%
\bibitem [{\citenamefont {Dassios}\ \emph {et~al.}(1994)\citenamefont
  {Dassios}, \citenamefont {Hadjinicolaou},\ and\ \citenamefont
  {Payatakes}}]{Dassios1994}%
  \BibitemOpen
  \bibfield  {author} {\bibinfo {author} {\bibfnamefont {G.}~\bibnamefont
  {Dassios}}, \bibinfo {author} {\bibfnamefont {M.}~\bibnamefont
  {Hadjinicolaou}}, \ and\ \bibinfo {author} {\bibfnamefont {A.~C.}\
  \bibnamefont {Payatakes}},\ }\href@noop {} {\bibfield  {journal} {\bibinfo
  {journal} {Q. Appl. Math.}\ }\textbf {\bibinfo {volume} {52}},\ \bibinfo
  {pages} {157} (\bibinfo {year} {1994})}\BibitemShut {NoStop}%
\bibitem [{\citenamefont {Salipante}\ and\ \citenamefont
  {Vlahovska}(2010)}]{Salipante2010}%
  \BibitemOpen
  \bibfield  {author} {\bibinfo {author} {\bibfnamefont {P.~F.}\ \bibnamefont
  {Salipante}}\ and\ \bibinfo {author} {\bibfnamefont {P.~M.}\ \bibnamefont
  {Vlahovska}},\ }\href@noop {} {\bibfield  {journal} {\bibinfo  {journal}
  {Phys. Fluids}\ }\textbf {\bibinfo {volume} {22}},\ \bibinfo {pages} {112110}
  (\bibinfo {year} {2010})}\BibitemShut {NoStop}%
\bibitem [{\citenamefont {Park}\ \emph {et~al.}(2009)\citenamefont {Park},
  \citenamefont {Ryu}, \citenamefont {Kim},\ and\ \citenamefont
  {Kang}}]{Parkand2009}%
  \BibitemOpen
  \bibfield  {author} {\bibinfo {author} {\bibfnamefont {J.~K.}\ \bibnamefont
  {Park}}, \bibinfo {author} {\bibfnamefont {J.~C.}\ \bibnamefont {Ryu}},
  \bibinfo {author} {\bibfnamefont {W.~K.}\ \bibnamefont {Kim}}, \ and\
  \bibinfo {author} {\bibfnamefont {K.~H.}\ \bibnamefont {Kang}},\ }\href@noop
  {} {\bibfield  {journal} {\bibinfo  {journal} {J. Phys. Chem. B}\ }\textbf
  {\bibinfo {volume} {113}},\ \bibinfo {pages} {12271} (\bibinfo {year}
  {2009})}\BibitemShut {NoStop}%
\bibitem [{\citenamefont {Kim}\ \emph {et~al.}(2011)\citenamefont {Kim},
  \citenamefont {Duprat}, \citenamefont {Tsai},\ and\ \citenamefont
  {Stone}}]{Kim2011}%
  \BibitemOpen
  \bibfield  {author} {\bibinfo {author} {\bibfnamefont {P.}~\bibnamefont
  {Kim}}, \bibinfo {author} {\bibfnamefont {C.}~\bibnamefont {Duprat}},
  \bibinfo {author} {\bibfnamefont {S.~S.~H.}\ \bibnamefont {Tsai}}, \ and\
  \bibinfo {author} {\bibfnamefont {H.~A.}\ \bibnamefont {Stone}},\ }\href@noop
  {} {\bibfield  {journal} {\bibinfo  {journal} {Phys. Rev. Lett.}\ }\textbf
  {\bibinfo {volume} {107}},\ \bibinfo {pages} {034502} (\bibinfo {year}
  {2011})}\BibitemShut {NoStop}%
\end{thebibliography}%

\end{document}